\documentclass[prl,twocolumn,amsmath,amssymb]{revtex4}

\usepackage{graphicx}
\usepackage{amsfonts}
\usepackage{amsmath}
\usepackage{amssymb}
\usepackage{color}
\usepackage{bm}
\usepackage{bbm}
\usepackage{enumerate}
\usepackage{amsfonts, amsmath, amsthm, amssymb} 
\usepackage{mathtools}
\usepackage{array}
\usepackage[colorlinks,bookmarks=false,citecolor=blue,linkcolor=red,urlcolor=blue]{hyperref}

\usepackage[usenames,dvipsnames,svgnames,table]{xcolor}
\graphicspath{figures} % Location of the graphics files

\usepackage[normalem]{ulem}

\newcommand{\old}[1]{\textcolor{Blue}{\sout{#1}}}

\renewcommand{\old}[1]{}

\begin{document}

\title{Thermal critical dynamics from equilibrium quantum fluctuations}

\author{Ir\'en\'ee Fr\'erot$^{1,2,3}$\footnote{Electronic address: \texttt{irenee.frerot@neel.cnrs.fr}},
Adam Ran\c con$^{4}$\footnote{Electronic address: 
\texttt{adam.rancon@univ-lille.fr}}
and Tommaso Roscilde$^{5}$\footnote{Electronic address: 
\texttt{tommaso.roscilde@ens-lyon.fr}}
 }
 
\affiliation{$^1$ ICFO - Institut de Ciencies Fotoniques, The Barcelona Institute of Science and Technology, 08860 Castelldefels (Barcelona), Spain}
\affiliation{$^2$ Max-Planck-Institut f\"ur Quantenoptik, Hans-Kopfermann-Stra{\ss}e 1, 85748 Garching, Germany}
\affiliation{$^3$ Univ Grenoble  Alpes, CNRS, Grenoble INP, Institut Néel, 38000 Grenoble, France}
 \affiliation{$^4$  Univ. Lille, CNRS UMR 8523 - PhLAM - Laboratoire des Lasers Atomes et Molécules, F-59000 Lille, France}
\affiliation{$^5$ Univ Lyon, Ens de Lyon, Univ Claude Bernard, CNRS, Laboratoire de Physique, F-69342 Lyon, France}
\date{\today}

%----------------------------------------------------------------------------------------
%	ABSTRACT
%----------------------------------------------------------------------------------------

\begin{abstract}
We show that quantum fluctuations display a singularity at thermal critical points, involving the dynamical $z$ exponent. Quantum fluctuations, captured by the quantum variance [I. Fr\'erot and T. Roscilde, Phys. Rev. B {\bf 94}, 075121 (2016)],
%We show that the dynamical critical exponent $z$, governing the critical dynamics close to a finite-temperature second-order phase transition, appears explicitly in the weak power-law singularity exhibited by the order-parameter \emph{quantum} fluctuations, captured, \emph{e.g.}, by the recently introduced quantum variance . As  
can be expressed via purely static quantities; this in turn allows us to extract the $z$ exponent related to the intrinsic Hamiltonian dynamics via \emph{equilibrium} unbiased numerical calculations, without invoking any effective classical model for the critical dynamics. These findings illustrate that, unlike classical systems, in quantum systems static and dynamic properties remain inextricably linked even at finite-temperature transitions, provided that one focuses on static quantities that do not bear any classical analog, namely on quantum fluctuations.    
\end{abstract}
 \maketitle

%----------------------------------------------------------------------------------------
%	INTRODUCTION
%----------------------------------------------------------------------------------------

Characterizing the dynamical behavior of systems close to symmetry-breaking phase transitions  \cite{tauber_book} has far-reaching implications, ranging from our understanding of the early Universe to the behavior of interacting systems in laboratory experiments \cite{Dziarmaga2010,delCampoZ2014,BeugnonN2017}. Indeed, close to a critical point, the fluctuations of the order parameter become correlated over a characteristic distance (the correlation length $\xi$) which can exceed arbitrarily the microscopic length (namely the distance between the elementary constituents); as a consequence, such fluctuations build up over a characteristic timescale (the relaxation time $t_c$) much larger than microscopic timescales. In particular, in the vicinity of a thermal critical point at temperature $T_c$, $\xi$ and $t_c$ are expected to be related to the temperature $T$ via
\begin{equation}
	t_c \sim \xi^{z} \sim  |T-T_c|^{-\nu z} ~,
	\label{eq_scaling_relation_tc_xi} 
\end{equation}
where $\nu$ governs the divergence of $\xi$, and $z$ is the dynamical exponent \cite{HohenbergH1977, tauber_book} governing the so-called critical slowing down of the order-parameter dynamics. A central goal of the study of critical phenomena is the evaluation of the various critical exponents for a given universality class. 

{
A fundamental paradigm in classical physics is that, in order to extract the dynamical exponent $z$ at a thermal phase transition, it is necessary to go beyond static thermodynamic observables, and to investigate instead the dynamics at criticality. 
Indeed, all static quantities can be obtained from the knowledge of the free energy $F = -k_B T \log Z$ (where $Z$ is the partition function) and of its derivatives; and, for classical systems, $F$ is completely independent of the dynamics. As a matter of fact, quantities such as position and momentum enter independently in the statistical sum over phase space, and, as an example, the partition function of a system of classical particles is fully independent of whether these particles have a ballistic dynamics, a stochastic dynamics, \emph{etc.}; therefore the dynamical exponent $z$ cannot be obtained from the knowledge of the free energy. On the other hand the $z$ exponent governs the singular behavior of dynamics, in accordance with dynamical scaling theory \cite{tauber_book}, which has been confirmed either directly by measuring the dynamical response functions in the vicinity of a critical point \cite{collins_book,fernandezetal1998,tsengetal2016,thesis_tseng}; indirectly via signatures of the Kibble-Zurek mechanism \cite{navonetal2015,ebadietal2021}; and in numerical simulations of the dynamics of (classical) microscopic models or classical field theories \cite{landauK1999,krechL1999,tsaiL2003, bergesetal2010,cheslereal2015,nandiT2019,nandiT2020}.

In the case of quantum systems, extracting \textit{ab initio} the $z$ exponent at thermal transitions for a given quantum Hamiltonian represents a notoriously difficult task. On the theory side, the effective coarse-grained dynamics to be studied can severely depend on the approximations made (\emph{e.g.} how collective modes are effectively included), see for instance \cite{saitoetal,morimatsuetal,mesterhazyetal}; in numerical simulations, computing the fully quantum real-time dynamics, or performing the analytical continuation of imaginary-time correlators \cite{JarrellG1996}, are prohibitive tasks for many-body quantum systems.  

Yet, in spite of the above cited difficulties to investigate their dynamics, quantum systems differ fundamentally from classical ones in that, in  quantum mechanics, static and dynamic properties remain inextricably intertwined. As a striking illustration, at zero-temperature (quantum) phase transitions the corresponding dynamical exponent does govern the scaling of the (free) energy \cite{sachdev_book}. The link between between statics and dynamics must remain true at finite-temperature transitions as well, given that the quantum-mechanical partition function is the trace of the imaginary-time evolution operator. In this work we probe this connection by considering  microscopic quantum models possessing a thermal phase transition, and whose order parameter is not a conserved quantity, namely it does not commute with the Hamiltonian. As a consequence, the order parameter has quantum fluctuations in addition to thermal fluctuations \cite{FrerotR2016,streltsovetal2017}. We show that these quantum fluctuations exhibit a weak singularity at thermal transitions, governed by a combination of critical exponents involving the dynamical $z$ exponent. 

Our prediction for the singularity of quantum fluctuations stems from the dynamical scaling hypothesis, the fluctuation-dissipation theorem and Kramers-Kronig relations; and it is rigorously verified using several exactly solvable quantum models of thermal phase transitions. As quantum fluctuations can be obtained from the partition function, this singularity establishes a fundamental link between the thermal critical dynamics of a quantum system and its statistical properties. Our results are not in contradiction with the notion that thermal criticality is fundamentally of classical origin, as the leading singular behavior of the order-parameter fluctuations coincides with that of the classical limit of the models of interest; yet the quantum fluctuations of the order parameter introduce a \emph{sub}-leading singular behavior (containing the exponent $z$) which can be analytically singled out, and which disappears in the classical limit.    
At the practical level, this insight allows one to extract the thermal $z$ exponent of microscopic quantum Hamiltonians without simulating the many-body dynamics at all. We demonstrate that our approach can be carried out successfully via numerically exact quantum Monte-Carlo (QMC) calculations on paradigmatic quantum spin models exhibiting thermal transitions. 
}

  To avoid any confusion, it should be stressed that the behavior of quantum fluctuations at thermal critical points, as studied in this paper, is completely independent of their behavior at (zero-temperature) quantum critical points \cite{sachdev_book,Haukeetal2016,FrerotR2018,FrerotR2019}, which are of no relevance to the present study. We also stress that the thermal singularity of quantum fluctuations is different from that of entanglement estimators, such as negativity, whose thermal critical behavior still lacks a general understanding \cite{Shermanetal2016,Waldetal2020,Luetal2020,Wuetal2020}.

\textit{Dynamical scaling hypothesis, and thermal singularity of the quantum variance.--}
The linear response of a system at thermal equilibrium to a weak time-dependent perturbation is characterized by its dynamical susceptibility \cite{forster_book, tauber_book}. If $O$ denotes the order parameter of a symmetry-breaking phase transition, and if a weak time-dependent perturbation in the form $-f(t) O$ is added to the Hamiltonian ${\cal H}$ of the system, the dynamical susceptibility is defined as $\chi_O(\omega) = \delta \langle O \rangle(\omega) / \delta f(\omega)|_{f=0}$, where $\langle O \rangle(\omega)$ and $f(\omega)$ are the Fourier transforms of $\langle O \rangle(t)$ and $f(t)$, respectively. Here, $\langle O \rangle(t) = Z^{-1} {\rm Tr}[O(t) \exp{(-\beta {\cal H}})]$ denotes an average over the equilibrium distribution at inverse temperature $\beta=(k_BT)^{-1}$, with $k_B$ the Boltzmann constant, and $O(t)=e^{i{\cal H}t/\hbar} O e^{-i{\cal H}t/\hbar}$ is the time-evolved operator in the Heisenberg picture. According to the dynamical scaling hypothesis \cite{HohenbergH1977, tauber_book}, close to the critical point, the singular part of the spectral function $\chi''_O(t)=\frac{1}{2\hbar}\langle [O(t),O(0)]\rangle$ (the imaginary part of $\chi_O(t)$) obeys in Fourier space the scaling form
%imaginary part of the dynamical susceptibility obeys the scaling form
\begin{equation}
	[\chi_O''(\omega)]_{\rm s} = [\chi^{(0)}_O]_{\rm s}~ g\left(\frac{\omega}{\omega_c}\right) ~,
	\label{eq_dynamical_scaling_hypothesis}
\end{equation}
where $\omega_c\sim t_c^{-1}\sim  |T-T_c|^{\nu z}$; $g(x)$ is a scaling function such that $\int dx ~g(x)/x = \pi$;  and $[\chi^{(0)}_O]_{\rm s} \sim |T-T_c|^{-\gamma}$ is the singular part of the static susceptibility of the order parameter with critical exponent $\gamma$. 

%In principle, the verification of the dynamical scaling hypothesis, and in particular the extraction of the dynamical exponent $z$, requires the full access to $\chi(\omega)$ at low frequency. However, we show in the following how the dynamical exponent can be extracted from purely static quantities.
%\textit{Weak singularity of the quantum variance across a critical point.--} 

In classical systems the variance of the order parameter ${\rm Var}(O) = \langle O^2 \rangle - \langle O \rangle^2$ is related to the susceptibility via the fluctuation-response relation ${\rm Var}(O) = k_B T \chi^{(0)}_O$, and therefore it exhibits the same power-law singularity.   
%Similarly to the static susceptibiliy, the variance of the order parameter $\langle \delta^2 O \rangle = \langle O^2 \rangle - \langle O \rangle^2$ diverges close to the critical point as
%\begin{equation}
%	\langle{\rm Var}(O) \approx k_B T \chi^{(0)}_O  \sim |T-T_c|^{-\gamma} ~,
%	\label{eq_scaling_chi_stat}
%\end{equation}   
%When quantum fluctuations are ignored, the equality ${\rm Var}(O) = k_B T\chi_O(0)$ holds in virtue of a fluctuation-response relation valid for classical systems. 
In quantum systems, the above relation holds only if $O$ is a conserved quantity, namely if $[{\cal H},O]=0$; otherwise quantum fluctuations are responsible for an extra contribution to the variance, the quantum variance (QV), ${\rm Var}_Q(O) ={\rm Var}(O) - k_B T\chi_O^{(0)} > 0$ \cite{FrerotR2016}. Our central result is that, at thermal criticality, the QV of the order parameter (along with a whole family of coherence measures to which it belongs) acquires a singular part scaling as
\begin{equation}
	[{\rm Var}_Q(O)]_{\rm s} \approx  A_1 |T-T_c|^{2\nu z-\gamma} + A_2 |T-T_c|^{4\nu z-\gamma} + ...
	\label{eq_scaling_QV}
\end{equation}
Given that $2\nu z-\gamma \ge 0$ for all the universality classes reported in the literature \cite{HohenbergH1977, tauber_book} this result has the immediate implication that the QV of the order parameter does \emph{not} diverge at the critical point, despite being the difference between two divergent quantities. This is consistent with the intuition that quantum fluctuations do not alter the long wavelength behavior of the system at thermal criticality -- in fact, their characteristic length scale, the quantum coherence length, remains finite even at $T_c$ \cite{MalpettiR2016}. At the same time, the weak singularity of the QV fully exposes the dynamical critical exponent, despite depending only on equilibrium fluctuations and the static response of the system.

\textit{Proof of the main result.--}
Eq. \eqref{eq_scaling_QV} follows directly from the dynamical scaling hypothesis (Eq. \eqref{eq_dynamical_scaling_hypothesis}). It is indeed a basic result of linear response theory that both ${\rm Var}(O)$ and $\chi_O^{(0)}$ can be expressed in terms of the imaginary part of the dynamical susceptibility \cite{tauber_book, forster_book}:
\begin{equation}
\begin{split}
	{\rm Var}(O) =& \hbar \int_{0}^{\infty} \frac{d\omega}{\pi} ~{\rm coth}(\beta \hbar \omega / 2) ~\chi_O''(\omega)~, \\
	\chi_O^{(0)} =& 2 \int_{0}^{\infty} \frac{d\omega}{\pi} ~\frac{\chi_O''(\omega)}{\omega}	~.
	\label{e.omegaintegrals}
\end{split}
\end{equation}
The first line is a consequence of the fluctuation-dissipation theorem \cite{callenW1951}, and the second of causality \cite{tauber_book, forster_book}.
Consequently, the QV admits the following expression 
\begin{equation}
	{\rm Var}_Q(O) = \hbar \int_{0}^{\infty} \frac{d\omega}{\pi} h_{\rm QV}(\beta \hbar \omega) \chi_O''(\omega)
	\label{eq_expression_QV_chi_sec}
\end{equation}
where $h_{\rm QV}(x) = {\rm coth}(x/2) - 2 / x$ filters out the low frequency modes in the integral (namely the modes such that $\hbar \omega \ll k_B T$). In contrast, the functions of which is it composed ($\coth(x/2)$ and $2/x$) both diverge at zero frequency. The dynamical scaling hypothesis, Eq.~\eqref{eq_dynamical_scaling_hypothesis}, suggests that the characteristic frequency singled out by $\chi_O''(\omega)$ (corresponding \emph{e.g.} to its maximum), moves to zero as $T\to T_c$, and therefore the critical singularity of both ${\rm Var}(O)$ and $\chi_O^{(0)}$ stems from the critical enhancement of the zero frequency contribution to the integrals Eq.~\eqref{e.omegaintegrals}. Similarly, the singular part of the quantum variance must also stem from the low-frequency part of the integral Eq.~\eqref{eq_expression_QV_chi_sec}. Assuming on other hand that $\chi_O''(\omega)$ vanishes as $\omega \to \infty$ faster than any power \footnote{Since $\int_{-\infty}^\infty d\omega \omega^n \chi''_O(\omega) \propto \partial_t^n \chi''_O(t)\big|_{t=0}$ and using the Heisenberg equation of motion $i\hbar\partial_t O(t)=[O(t),H]$, we observe that these integrals are proportional to averages of nested commutators $\langle[[[O,\ldots],H],H]\rangle$ which are expected to be finite on physical ground.}
we may then Taylor expand $h_{\rm QV}(x)$ at low frequency as $h_{\rm QV}(x) = x/6 - x^3/360 + ... $, to obtain  		
	\begin{equation}
	[{\rm Var}_Q(O)]_{\rm s} \approx [\chi_O^{(0)}]_{\rm s} \left [ \frac{(\hbar \omega_c)^2}{k_B T} {\cal I}_1 + \frac{(\hbar \omega_c)^4}{(k_B T)^3}  {\cal I}_2  + ...\right]~,
	\label{eq_expression_QV_sing}
	\end{equation}	 
		where ${\cal I}_1 = (6\pi)^{-1}\int_0^\infty dx ~ x g(x)$, ${\cal I}_2=- (360\pi)^{-1}\int_0^\infty dx ~ x^3~ g(x)$, etc. Given the critical behavior of $[\chi_O^{(0)}]_{\rm s}$ and $\omega_c$, we obtain Eq. \eqref{eq_scaling_QV}.

\textit{Extension to asymmetry measures.--}
Eq. \eqref{eq_expression_QV_chi_sec} is very similar to an expression derived for the quantum Fisher information (QFI) \citep{Haukeetal2016} (with $h_{\rm QFI}(x) = {\rm tanh}(x/2)$), and in fact, both the QV and the QFI belong to a larger family of so-called quantum coherence (or ``asymmetry'') estimators \cite{streltsovetal2017}, all admitting an analogous expression in term of $\chi''$ (see the Supplementary Material -- SM \cite{SM} for further details). Most importantly, for all the coherence measures of this family the ``quantum filter'' $h(x)$ appearing in the frequency integrals of the dynamical susceptibility has an odd parity, and therefore is linear at low frequency \cite{Haukeetal2016,SM}. The latter property, together with the dynamical scaling hypothesis (Eq. \eqref{eq_dynamical_scaling_hypothesis}), are the only requirements leading to Eq. \eqref{eq_scaling_QV}; hence our result immediately applies to all of them. In the specific case of the QFI, our result rectifies a statement of Ref.~\citep{Haukeetal2016} on the absence of thermal singularities \cite{SM}.

\noindent \textit{Two exactly-solvable models.}
We first illustrate our findings with two quadratic models which belong to the same static universality class, yet having different $z$ exponents (we also treat the case of a quantum Ising model with infinite range interaction in \cite{SM}). The first one is the so-called quantum spherical model (QSM) on a $d$-dimensional lattice \cite{tuW1994,nieuwenhuizen1995,Vojta1996}, defined by the Hamiltonian ${\cal H}_{\rm QSM}= \frac{g}{2}\sum_i  P_i^2+\frac{1}{2g}\sum_{i,j} U(i,j)  X_i  X_j +\lambda\left(\sum_i X_i^2-\frac{N}{4}\right)$ , $[X_i,P_i]=i\hbar \delta_{ij}$, and $U(i,j)$ defines the interaction. The second model describes a Bose-Einstein condensation (BEC) transition \cite{guntonB1968,Jakubczyk2013,Jakubczyk2018}, and is defined by ${\cal H}_{\rm BEC}=-\mu N_B+\sum_{\bf k} (\epsilon_k+\mu) a^\dag_{\bf k} a_{\bf k}$, $[a_{\bf k},a_{\bf l}]=\delta_{\bf kl}$. The Lagrange parameter $\lambda$ (resp.~$\mu$) imposes the constraint $\sum_i \langle X_i^2 \rangle=N/4$ (resp.~$\sum_{\bf k} \langle a^\dag_{\bf k} a_{\bf k} \rangle=N_B$), with $N$ the number of sites (resp.~$N_B$ the number of bosons). In both models, we assume that at small momenta $\epsilon_k\sim U_k\sim k^x$, with $d<x<2d$  ($x=2$ corresponds to short-range interactions in the QSM). Both models have a phase transition at a critical temperature $T_c$, and belong to the universality class of the spherical model \cite{Berlin1952,joyce1966}, with exponents $\nu=\frac{1}{d-x}$ and $\gamma=\frac{x}{d-x}$ \cite{joyce1966,Vojta1996,Jakubczyk2018}, so that $\lambda, \mu \propto |T-T_c|^{\frac{x}{d-x}}$. In the SM \cite{SM}, we show that for the QSM, the dynamical exponent is $z_{\rm QSM}=x/2$, and that the QV of the order parameter ($O_{\rm QSM}=N^{-1/2}\sum_i X_i$) is ${\rm Var}_Q(O_{\rm QSM})= \frac{\hbar g }{2\omega_0} h_{\rm QV}(\beta \hbar \omega_0)$, with critical frequency $\omega_0=\sqrt{2g \lambda}$. For the bosons ($O_{\rm BEC} = a_{\bf 0} + a_{\bf 0}^\dagger$), we find $z_{\rm BEC}=x$ and ${\rm Var}_Q(O_{\rm BEC})= h_{\rm QV}(\beta \mu)$. Close to criticality, we therefore obtain:
\begin{equation}
\begin{split}
{\rm Var}_Q(O_{\rm QSM})&= \frac{\hbar^2 g}{12 k_B T }-\frac{\hbar^4 g^2\lambda}{720 (k_B T)^3}+\ldots\, ,\\
{\rm Var}_Q(O_{\rm BEC})&= \frac{\mu}{6k_B T }-\frac{\mu^3}{360 (k_B T)^3}+\ldots\, .
\end{split}
\end{equation}
The singular part of the QV in the QSM model stems from the (negative) ${\cal I}_2$ contribution to Eq.~\eqref{eq_expression_QV_sing}, scaling as $|T - T_c|^{4\nu z - \gamma}$, which is consistent with $z_{\rm QSM}=x/2$.
Remarkably, the QV vanishes at the BEC transition, as well as in the whole BEC phase \footnote{Strictly speaking, the QV is exponentially small in the system size.}. The singular contribution scales as $|T-T_c|^{2\nu z - \gamma}$ [Eq.~\eqref{eq_scaling_QV}], which is consistent with $z_{\rm BEC}=x$.\\

\begin{figure*}
	\centering
		\includegraphics[width=1.\linewidth]{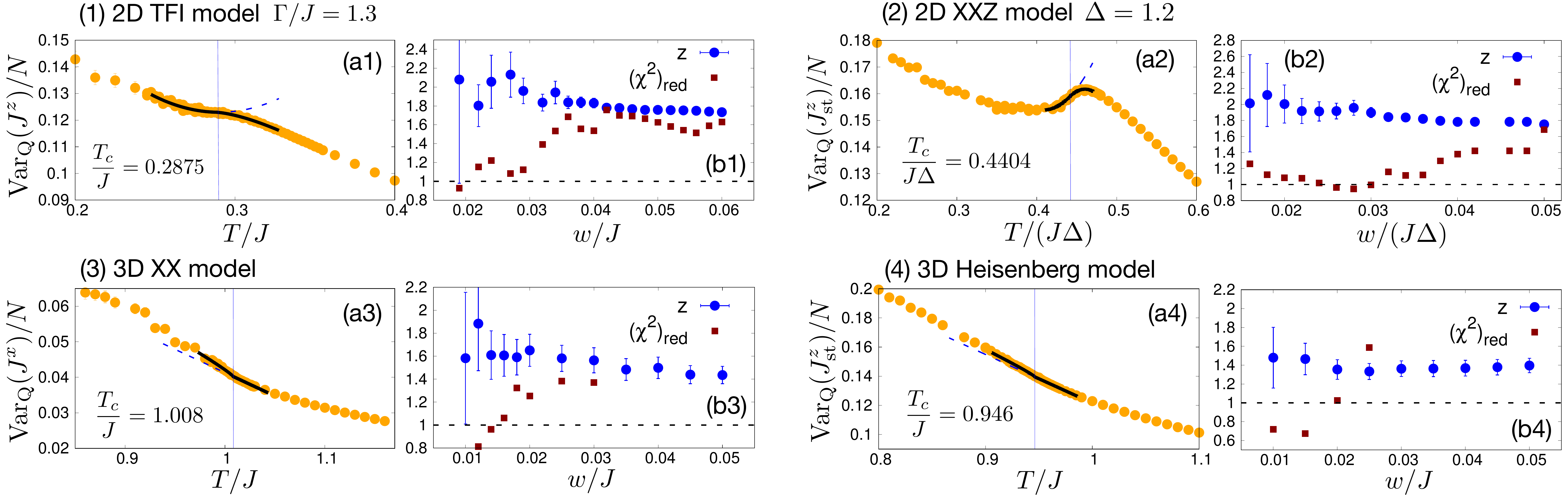}
		\caption{Thermal singularity of the quantum variance from QMC results, and fitted $z$ exponents. The (a1-a4) panels show the quantum variance of the order parameter close to the transition, together with a representative fit (black line) and the fitted regular part (blue dashed line); the vertical line marks the transition point. The (b1-b4) panels show the resulting fitted $z$ exponent as as function of the fitting window width $w$ around $T_c$, along with the reduced $\chi^2$ (the dashed line marks the unity threshold).  (a1-b1):  2$d$ TFIM ($\Gamma/J = 1.3$ - lattice size $L=64$); (a2-b2) 2$d$ antiferromagnetic XXZ model ($\Delta = 1.2$, $L=64$); (a3-b3) 3D XX model ($L=28$); (a4-b4) 3D Heisenberg model ($L=28$).}
		\label{f.zQMC}
	\end{figure*}

\textit{Extraction of the dynamical exponent using QMC.}
 The QV can be calculated for \emph{any} quantum model whose thermodynamics can be calculated efficiently \cite{FrerotR2016}, opening the route for a systematic calculation of the $z$ exponent in a large class of quantum many-body systems. We illustrate this possibility by focusing on four paradigmatic models exhibiting a finite-temperature transition, defined on $d$-dimensional (hyper-)cubic lattices of size $L^d$: (1) the 2$d$ ferromagnetic transverse-field Ising model (TFIM)  
 ${\cal H}_{\rm TFIM} = -J \sum_{\langle ij \rangle} S_i^z S_j^z - \Gamma \sum_i S_i^x$ with $\Gamma/J = 1.3$; and the XXZ model ${\cal H}_{\rm XXZ} = - J \sum_{\langle ij \rangle}  \left ( S_i^x S_j^x + S_i^y S_j^y - \Delta S_i^z S_j^z \right )$ (with $J<0$) in the three following cases: (2) 2$d$ easy-axis model ($\Delta = 1.2$); (3) 3$d$ XX model ($\Delta = 0$); and (4) 3$d$ Heisenberg model $\Delta = 1$) \footnote{The standard Heisenberg model has equal (positive) sign for the couplings of all three spin components; yet on bipartite lattices it can be mapped onto the XXZ model -- considered here -- with ferromagnetic couplings for the $x$ and $y$ components.}. For all models $\langle ij \rangle$ are nearest-neighbor pairs on a $d$-dimensional  lattice. All four models possess a finite-temperature transition. For models 1 and 2, the transition belongs to the 2$d$ Ising universality class. However in model 2 the magnetization along $z$ is conserved: it represents a diffusive mode which could potentially couple to the order parameter and alter the $z$ exponent with respect to model 1 \cite{tauber_book}. Model 3 is representative of the 3$d$ XY universality class, and model 4 of the 3$d$ Heisenberg class.
 We investigate all models using Stochastic Series Expansions quantum Monte Carlo \cite{SyljuasenS2002} which allows us to reconstruct the QV (as already shown in Refs.~\cite{FrerotR2016, FrerotR2018, FrerotR2019}). The system sizes we analyze ($L=64$ in $d=2$; $L=28$ in $d=3$) are not the largest ones we can simulate, but the QV exhibits very little scaling beyond these sizes \cite{SM}, while its precision degrades significantly in the ordered phase, as the QV (per spin) is the non-diverging difference between two divergent quantities. For all the models, $T_c$ is estimated by a scaling analysis of the full variance of the order parameter (scaling as $L^{\gamma/\nu}$ at the transition point). 
 
The QV per spin of the order parameter (uniform magnetization $O = J^z$ for the 2$d$ TFIM, $O = J^x$ for the 3$d$ XX model, and staggered magnetization $O=J^z_{\rm st}$ for the 2$d$ XXZ model and the 3$d$ Heisenberg model) shows a clear anomaly at the phase transition (Fig.~\ref{f.zQMC}). It is then fitted as:
\begin{eqnarray}
{\rm Var}_Q(O)(T)  &=&  a_0 + a_1 T + a_2 T^2 \nonumber \\
&+&   \left [ A_+~  \theta(\delta) +  A_- \theta(-\delta) \right ]  |\delta|^{2\nu z-\gamma} ~.
\label{e.fit}
\end{eqnarray}
where $\delta=T-T_c$. The first line is a parabolic fit to the regular part, while the second line is a fit to the dominant term of the singular part (the critical exponents $\nu$ and $\gamma$ are known for each universality class \cite{PelissettoV2002}). 
In principle we have 6 fitting parameters ($a_0, a_1, a_2, A_+, A_-, z$), which are nonetheless further reduced: a) for the $d=2$ models (1 and 2), we set $A_-=0$, as suggested by the smallness of $A_-$ when treated as a free parameter; b) for the $d=3$ models, we set $A_+=0$ (for the same reason as above) as well as $a_2=0$. A subtle aspect of the fits is the discrimination of the singular part from the regular one, since both are non-divergent. This is particularly true for the models 1 and 2, for which an alternative fitting analysis is presented in the SM \cite{SM}.

We perform our fits to Eq.~\eqref{e.fit} over windows of variable width $[T_c-w,T_c+w]$ around the critical point. 
 Fig.~\ref{f.zQMC} shows the results: the fit quality is always very good for all four models, with $(\chi^2)_{\rm red}$ ($\chi^2$ per degree of freedom) systematically reaching values around 1 upon shrinking the fitting window down to the relevant critical region. As for our final estimates of $z$, we retain the values at which $(\chi^2)_{\rm red} \approx 1$ and for which the fitted value has converged upon reducing $w$ (within the error bar): (1) $z = 1.95(15)$; (2) $z=1.95(10)$; (3) $z = 1.61(15)$; (4) $z=1.36(10)$.  Models 1 and 2 ($2d$ Ising universality class) could potentially be captured by Model C ($z=2$) \cite{HohenbergH1977,tauber_book}. The alternative fitting strategy presented in the SM \cite{SM} gives $z=1.88(10)$ for model 2, compatible with $z=1.95(10)$ obtained above, while it gives $z\simeq 1.65(5)$ for model 1, closer to experimental results on quasi-2$d$ magnets ($z\approx 1.6-1.8$) \cite{Buccietal1974, Hutchingsetal1982, Slivkaetal1984,tsengetal2016,thesis_tseng}. For model 3, our estimate is compatible with that obtained by Ref.~\cite{krechL1999}  ($z=1.62$) via a dynamical simulation of the classical 3d XY model. As for model 4, our estimate is slightly lower but compatible with $z=1.49(3)$ obtained via dynamical simulations of classical Heisenberg antiferromagnets \cite{Bunkeretal1996,landauK1999,tsaiL2003}, and with the experimental estimate from neutron scattering studies of quantum Heisenberg antiferromagnets \cite{Tucciaroneetal1971, Coldeaetal1998}.
  
\textit{Conclusions.--}
We have shown that the dynamical exponent $z$, governing the critical slowing down of the dynamics close to a second-order thermal phase transition of a quantum system, manifests itself in a weak singularity of the quantum fluctuations of the order parameter. This general result was illustrated by two exactly solvable models with the same thermodynamic criticality but different critical dynamics; by a quantum spin model with infinite-range interactions amenable to exact diagonalization for large system sizes (as discussed in the Supplemental Material \cite{SM}); and exploited to extract the exponent $z$ in four quantum spin models in $d=2$ and $d=3$ from unbiased QMC data. Our scheme gives access to the $z$ exponent associated with the \emph{intrinsic} Hamiltonian dynamics (in the absence of any external bath) without the need to simulate the real-time quantum dynamics itself (which is a prohibitive numerical task); and the $z$ exponent can be extracted from numerical (\emph{e.g.} QMC) data without the need for analytic continuation (which is also a very demanding task). Here we have focused on the singularity of quantum fluctuations of the order parameter, but other quantities are expected to display similar singularities exposing the $z$ exponent, offering alternative strategies for its numerical evaluation. Therefore our approach opens a way to the calculation of dynamical critical exponents for the Hamiltonian dynamics of a large class of quantum many-body models -- something which is of extreme importance in the light of the recent generation of experiments addressing the dynamics of closed quantum systems close to critical points \cite{Chomazetal2015,Braunetal2015,navonetal2015,BeugnonN2017}. Moreover, the ability of inelastic neutron scattering experiments to reconstruct quantum coherence estimators \cite{Mathewetal2020, Scheietal2021,Laurelletal2021, Scheieetal2021b} along with dynamical critical scaling \cite{collins_book} could lead to a direct test of our predictions within the same experimental platform. Finally, at a fundamental level, the precise connection between microscopic quantum models (as studied in this work) and effective classical theories \cite{HohenbergH1977,tauber_book,krechL1999,tsaiL2003,bergesetal2010} is a fascinating topic for future investigations.

\begin{acknowledgements}
\textit{Acknowledgments.} We are grateful to V. Alba, J. Dalibard, N. Laflorencie, N. Dupuis, and M. Lewenstein for stimulating discussions.
IF and TR acknowledge support by ANR (``ArtiQ" project, ``EELS" project) and QuantERA (``MAQS" project). IF acknowledges support by  the Government of Spain (FIS2020-TRANQI and Severo Ochoa CEX2019-000910-S), Fundació Cellex, Fundació Mir-Puig, Generalitat de Catalunya (CERCA, AGAUR SGR 1381 and QuantumCAT) the Agence Nationale de la Recherche (Qu-DICE project ANR-PRC-CES47), the John Templeton Foundation (Grant No. 61835).
AR is supported by the Research Grants QRITiC I-SITE ULNE/ ANR-16-IDEX-0004 ULNE. All authors acknowledge support of CNRS (PEPS project "EnIQMa"). 
\end{acknowledgements}

\bibliography{biblio_critical_dynamics}

\newpage
\appendix
\begin{center}
{\Large{\textbf{Supplemental Material}}}
\end{center}

\section{Singularity of a family of quantum coherence estimators.}
In this appendix, we show that a general family of quantum coherence estimators, to which both the quantum variance (QV) and the quantum Fisher information (QFI) belong, have an analogous expression in terms of the dynamical susceptibility. As a consequence, they all display a singularity of the same form at critical points. This appendix is a summary of a longer discussion presented in ref.~\citep[Sections 1.5-1.6]{thesis_frerot}. The relevant family of coherence estimators, quantifying the coherence of a quantum state $\rho$ with respect to an observable ${\cal O}$, can be written as:
\begin{equation}
    C_f(\rho, {\cal O}) = \frac{f(0)}{2} \sum_{i \neq j} 
    \frac{(p_i - p_j)^2}{p_i f(p_j / p_i)}|\langle i|{\cal O}|j \rangle|^2 ~,
    \label{eq_def_Cf}
\end{equation}
where $|i\rangle$ is the eigenvector of $\rho$ corresponding to eigenvalue $p_i$. Here, $f$ is a function obeying certain mathematical properties \cite{petz1996} ensuring that $C_f$ is a valid coherence estimator, among which $f(x)\ge 0$; $f(x)=xf(1/x)$ and $f(1)=1$. For pure states, it can be verified that $C_f(|\psi\rangle\langle \psi|, {\cal O}) = \langle \psi|{\cal O}^2|\psi\rangle - \langle\psi|{\cal O}|\psi\rangle^2$ is the variance; in general $C_f$ is smaller than the variance, and obey several mathematical properties which make it a well-defined estimator of quantum coherence \cite{thesis_frerot} (or, according to an alternative terminology, of ``asymmetry'' \cite{streltsovetal2017}). To this general family of coherence estimators belong the QFI (for which $f(x)=F(x)=\frac{1+x}{2}$), the QV ($f_{\rm QV}(x) = \frac{(1-x)^2/6}{1+x+2(1-x)/\log x}$), and the Wigner-Yanase-Dyson skew information, defined as $I_\alpha={\rm Tr}(\rho^{1-\alpha} {\cal O} \rho^\alpha {\cal O})$ ($f_\alpha(x)=\frac{\alpha(1-\alpha)(1-x)^2}{1+x-x^\alpha - x^{1-\alpha}}$). The fact that all coherence estimators $C_f$ behave in a qualitatively similar fashion across the phase diagram of a many-body system descends from the fundamental inequality \cite{gibiliscoetal2007}: $C_f \le C_F \le \frac{1}{2f(0)}C_f$, implying in particular that all $C_f$ have the same scaling behavior. Most importantly, for thermal states, straightforward manipulations and basic results of linear response theory allow to establish the following expression \cite{thesis_frerot}:
\begin{equation}
    C_f(\rho, {\cal O}) = \hbar \int_0^{\infty} \frac{d\omega}{\pi} h_f(\beta \hbar \omega) \chi''_{\cal O}(\omega) ~,
    \label{eq_Cf_Chi_sec}
\end{equation}
where $\beta$ is the inverse temperature, and with the ``quantum filter'':
\begin{equation}
h_f(x) = f(0) \frac{1-e^{-x}}{f(e^{-x})} ~.
\label{eq_Cf_filter}
\end{equation}
Eqs.~\eqref{eq_Cf_Chi_sec}-\eqref{eq_Cf_filter} extend the result for the QFI \cite{Haukeetal2016} and the QV (main text of this paper) to the complete family of coherence estimators defined via Eq.~\eqref{eq_def_Cf}. At small frequency ($\hbar \omega \ll k_BT$), one finds $h_f(x) \approx xf(0)$, namely $h_f$ is linear. Furthermore, the property $f(x)=xf(1/x)$ implies that $h_f(x)$ is an odd function, leading to an expression similar to Eq.~(6) of the main text. In conclusion: for all allowed $f$ functions \cite{petz1996}, $C_f(\rho, {\cal O})$ displays a singularity of the same form as that of the QV at critical transitions.

\section{Quantum variance of the quantum spherical model}

The  quantum spherical model (QSM) \cite{tuW1994,nieuwenhuizen1995,Vojta1996} is a quantum generalization of the spherical model \cite{Berlin1952,joyce1966}, and is defined in terms quantum operators $ X_i$ on a $d$-dimensional lattice with $N$ sites, with a mean-field constraint $\sum_i  \langle X_i^2\rangle=\frac{N}{4}$. The Hamiltonian is
\begin{equation}
 H_{\rm QSM} = \frac{g}{2}\sum_i  P_i^2+\frac{1}{2g}\sum_{i,j} U(i,j)  X_i  X_j +\lambda\left(\sum_i X_i^2-\frac{N}{4}\right),
\end{equation}
with $ P_i$ the canonical momentum $[X_i, P_j]=i\hbar\delta_{ij}$ (all other commutators vanish), $\lambda$ is a Lagrange multiplier, and $U(i,j)$ is the two-body interaction. The order parameter is $O := N^{-1/2}\sum_i X_i$. The QSM is essentially equivalent to the quantum non-linear sigma model with $O(N)$ symmetry in the large $N$ limit \cite{sachdev_book}.

 Assuming that $U(i,j)$ is invariant under translations, and that its Fourier transform $U_k\sim k^x$ at small momenta ($x<d<2x$, and $x=2$ for short-range interaction), the free energy is given by \cite{Vojta1996}
\begin{equation}
f_{\rm QSM}=-\frac{\lambda}{4}+\frac{k_B T}{N}\sum_{\bf k}\ln\left(2\sinh\left(\frac{\hbar\omega_k}{2 k_B T}\right)\right),
\end{equation}
with $\omega_k^2=U_k+2g\lambda$. The mean-field constraint is imposed by
\begin{equation}
\frac{\partial f_R}{\partial \lambda}=0.
\label{eq_dfdmu}
\end{equation}
Solving this equation for the Lagrange multiplier gives $\lambda(g,T)$. For a given $g$, there is a finite-temperature phase transition at a critical temperature $T_c(g)$, defined by $\lambda(g,T_c(g))=0^+$. One finds $\lambda\propto \delta^{\frac{x}{d-x}}$ and $\omega_k^2\simeq k^x+2g \lambda$ at small momenta, with $\delta=|T-T_c|\ll 1$, which implies that the static critical exponents are those of the spherical model \cite{joyce1966}, namely $\nu=\frac{1}{d-x}$ and $\eta=0$ (and in particular $\gamma=\frac{x}{d-x}$) \cite{Vojta1996}.

The finite-momentum imaginary part of the dynamical susceptibility is computed from
\begin{equation}
\chi_{O}''(k,t)=\frac{1}{2\hbar N}\left\langle [ X_k(t), X_{-k}(0)]\right\rangle,
\end{equation}
which gives
\begin{equation}
\chi_O''(k,\omega)=\frac{\pi g}{2\omega_k}\left(\delta(\omega-\omega_k)-\delta(\omega+\omega_k)\right).
\end{equation}
At zero momentum, we get
\begin{equation}
\chi_{O}''(\omega)=\frac{\pi }{4 \lambda}\left(\delta\left(\frac{\omega}{\sqrt{2g\lambda}}-1\right)-\delta\left(\frac{\omega}{\sqrt{2g\lambda}}+1\right)\right),
\end{equation}
which obeys the scaling law $\chi_O''(\omega,\delta)\propto \delta^{-\gamma} f(\omega \delta^{-\frac{x}{2(d-x)}})$, from which we read $z_{\rm QSM}=x/2$.

A straightforward calculation gives the variance as ${\rm Var}(O)=N^{-1}\lim_{k\to 0}\langle X_k X_{-k}\rangle=2g \lim_{k\to 0}\frac{\partial f_R}{\partial U_k}$:
\begin{equation}
{\rm Var}(O)=\frac{\hbar g}{2\omega_0}\coth\left(\frac{\hbar \omega_0}{2k_B T}\right),
\end{equation}
and the static susceptibility:
\begin{equation}
\chi^{(0)}_{O}=\frac{g}{\omega_0^2}.
\end{equation}
The quantum variance thus reads
\begin{equation}
\begin{split}
{\rm Var}_Q(O)&=\frac{\hbar g}{2\omega_0} h_{\rm QV}(\beta \hbar \omega_0),\\
&\simeq \frac{\hbar^2 g}{12 k_B T }-\frac{\hbar^4 g^2\lambda}{720 (k_B T)^3}+\ldots\, .
\end{split}
\end{equation}
Comparing with Eq.~(6) of the main text, we see that the scaling of the quantum variance is consistent with $z_{\rm QSM}=x/2$, using $2\nu z_{\rm QSM}-\gamma=0$ and $4\nu z_{\rm QSM}-\gamma=x/(d-x)$.

\section{Quantum variance at a Bose-Einstein condensation with mean-field interaction}

A simple description of Bose-Einstein condensation (BEC) is obtained from a model of bosons with mean-field (all-to-all) interaction \cite{guntonB1968,Jakubczyk2013}. Alternatively, one can study free bosons with a constraint on the mean density $\frac1N\sum_i  \langle n_i\rangle =n_B$, with $n_i$ the number operator on the site $i$ and $n_B$ is the density of bosons. In momentum space, the Hamiltonian is given by
\begin{equation}
H_B=- N\mu n_B+\sum_{\bf k} E_k a^\dag_{\bf k} a_{\bf k},
\end{equation}
where $E_k=\epsilon_k +\mu$, $\mu$ is an effective chemical potential that plays the role of a Lagrange multiplier to enforce the constraint. We assume that the bosons have a modified dispersion $\epsilon_k\propto k^x$. The creation and annihilation operator satisfy the canonical commutation relations $[a_{\bf k},a_{\bf l}]=\delta_{\bf kl}$. 

The free energy of the bosons is given by
\begin{equation}
f_B=-\mu n_B-\frac{1}{\beta N}\sum_{\bf k}\ln\left(1-e^{-\beta E_k}\right),
\end{equation}
while the constraint reads
\begin{equation}
\frac{1}{N}\sum_{\bf k}\frac{1}{e^{\beta E_k}-1}=n_B.
\end{equation}
The transition happens at a critical temperature $T_c(n_B)$ such that $\mu=0^+$. A standard analysis gives $\mu\propto \delta^{\frac{x}{d-x}}$ \cite{Jakubczyk2018}, and thus the static critical exponent are those of the spherical model, namely $\nu=\frac{1}{d-x}$ and $\eta=0$.

An observable playing the role of an order parameter for the BEC transition is the field quadrature $q_{\bf k}=a_{\bf k}+a^\dag_{\bf k}$ at zero momentum. The imaginary part of its dynamical susceptibility is given by
\begin{equation}
\chi_q''(k,\omega)=\frac{\pi}{\hbar}\left(\delta(\omega-E_k)-\delta(\omega+E_k)\right),
\end{equation}
from which we read $\omega_c=\mu/\hbar$ and thus $z_{\rm BEC}=x$, which is twice that of the quantum spherical model. This is in agreement with the critical behavior of the quantum variance of $q_0$,
\begin{equation}
\begin{split}
{\rm Var}_Q(q)&=h_{\rm QV}(\beta \mu),\\
&\simeq \frac{\mu}{6k_B T }-\frac{\mu^3}{360 (k_B T)^3}+\ldots,
\end{split}
\end{equation}
since $\mu\propto  \delta^{\frac{x}{d-x}}=\delta^{2\nu z-\gamma}$ and $\mu^3\propto  \delta^{3\frac{x}{d-x}}=\delta^{4\nu z-\gamma}$.

\section{Ising model with infinite-range interactions}
In this section, we compute the quantum variance (QV) and quantum Fisher information (QFI) at the thermal phase transition of the $s=1/2$ transverse-field Ising model with infinite-range interactions:
\begin{equation}
{\cal H} = -\frac{1}{N} (J^z)^2 - g J^x ~,
\label{eq_Ising_LMG}
\end{equation}
where $J^a = \sum_{i=1}^N \sigma_i^a / 2$, with $\sigma_i^a$ the Pauli matrices ($a=x,y,z$). Upon varying $g$ from $0$ to $1$, this model displays a continuous line of second-order phase transitions at $T_c=g / \ln \left(\frac{1 + g}{1 - g} \right)$ \cite{dasetal2006} with mean-field exponents ($\nu=1/2,\gamma=1$). This represents another example of a quantum model for which the quantum coherence measures of interest (QV and QFI) can be computed exactly at any temperature. As we shall see in this section, these computations confirm our general result concerning the singular behavior of these quantities at a thermal transition, involving the dynamical $z$ exponent. Regarding the QFI, the present computation rectifies the results of Ref.~\cite{Haukeetal2016} which reported the absence of any singularity at the transition (Fig.~3 of Ref.~\cite{Haukeetal2016}).

First, we show that $z=1$ at the thermal transition; then, we investigate the singularity of the QV and the QFI at the thermal transition and relate it to the $z$ exponent.

\begin{figure}
	\centering
		\includegraphics[width=1.0\linewidth]{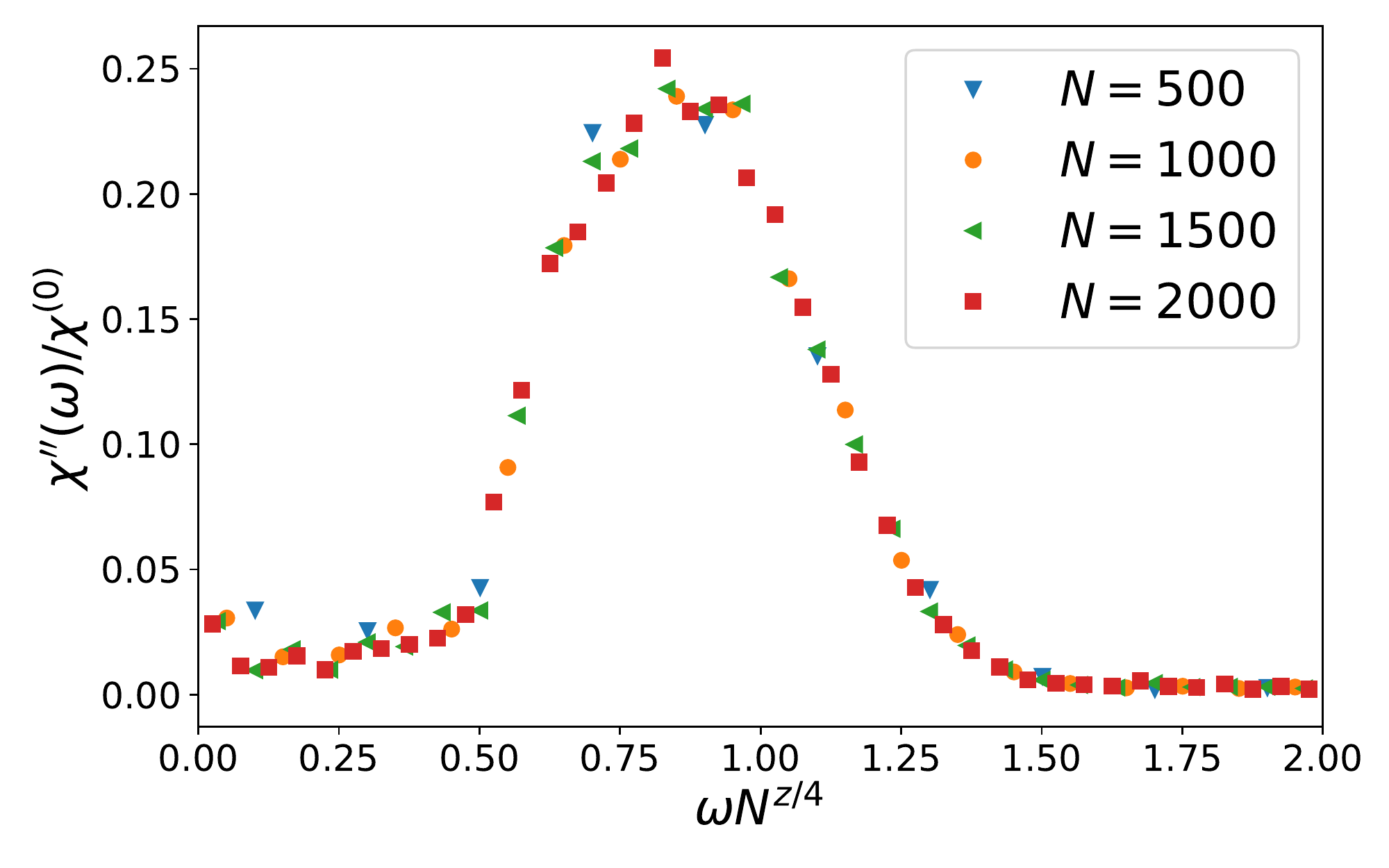}
		\caption{Dynamical susceptibility for the Ising model with infinite-range interactions ($g=0.8,T=T_c\approx 0.3641$). The imaginary part of the dynamical susceptibility $\chi''(\omega)$ of $J^z$ is rescaled to the static susceptibility $\chi^{(0)}$. It is plotted as a function of $\omega N^{z/4}$ with $z=1$. 
		}
		\label{fig_chi_sec_Ising_all_to_all}
	\end{figure}

\textit{Dynamical exponent.}
 As the total spin $(\bm J)^2$ commutes with the Hamiltonian, the latter can be diagonalized sector by sector, taking into account the degeneracy of each sector. The largest dimension ($N+1$) corresponds to a total spin $J=N/2$, which sets the maximal number of spins reachable in a numerical exact diagonalization. The dynamical susceptibility $\chi''(\omega)$  of the order parameter $J^z$ is given by $\chi''(\omega) = (1 - e^{-\beta \hbar \omega})S(\omega) / 2$, with $S(\omega)$ the dynamical structure factor \cite{callenW1951}:
\begin{eqnarray}
	S(\omega) = Z^{-1}\sum_{J=0}^{N/2} D(J) && \sum_{k,k'=-J}^{J} e^{-\beta E(J, k)}
	|\langle J,k | J^z | J, k' \rangle |^2 \times \nonumber \\
	&& 	\delta(\omega - [E(J,k') - E(J,k)]/\hbar)
	\label{eq_S_omega_Ising_all_to_all}
\end{eqnarray}
where $E(J,k)$ and $|J,k\rangle$ are the energy and eigenstates in a sector of fixed $J$. The factor $D(J)= {N \choose N/2-J} - {N \choose N/2-J-1}$ accounts for the degeneracy of each $J$ sector \citep{arecchietal1972}. In Fig.~\ref{fig_chi_sec_Ising_all_to_all}, we plot $\chi''(\omega) / \chi^{(0)}$ at the critical temperature for $g=0.8$, for different system sizes ($N=500,1000,1500,2000$). We computed $S(\omega)$ using Eq.~\eqref{eq_S_omega_Ising_all_to_all}, with frequency bins of width $100 / N^{5/4}$.

For a $d$-dimensional system of linear size $L$, the characteristic frequency $\omega_c$ scales according to $\omega_c(L) \sim L^{-z}$ with $z$ the dynamical exponent. In the case of infinite-range systems, $L$ is replaced by $N^{1/d}$ with $d$ the upper critical dimension ($d=4$ in the present case) \cite{botetetal1982}. In Fig.~\ref{fig_chi_sec_Ising_all_to_all}, the frequency is therefore rescaled to $\omega_c(N) \sim N^{-z/d}$, with the choice $z=1$ providing excellent collapse of the data for various system sizes. In summary, Fig.~\ref{fig_chi_sec_Ising_all_to_all} shows that the dynamical exponent is $z=1$ at the thermal transition. Interestingly, the value $z=1$ differs from the mean-field prediction $z=2$ for the critical dynamics of a non-conserved order parameter (model A) \cite{HohenbergH1977,tauber_book}. However, $z=1$ is in agreement with both the quantum O(1) model in a gaussian approximation \cite{thesis_frerot}; and with the quantum spherical model, which has the same symmetries as the Ising model, and which for $x=2$ has the same (mean-field) static critical exponents. \\

\begin{figure}
	\centering
		\includegraphics[width=1.0\linewidth]{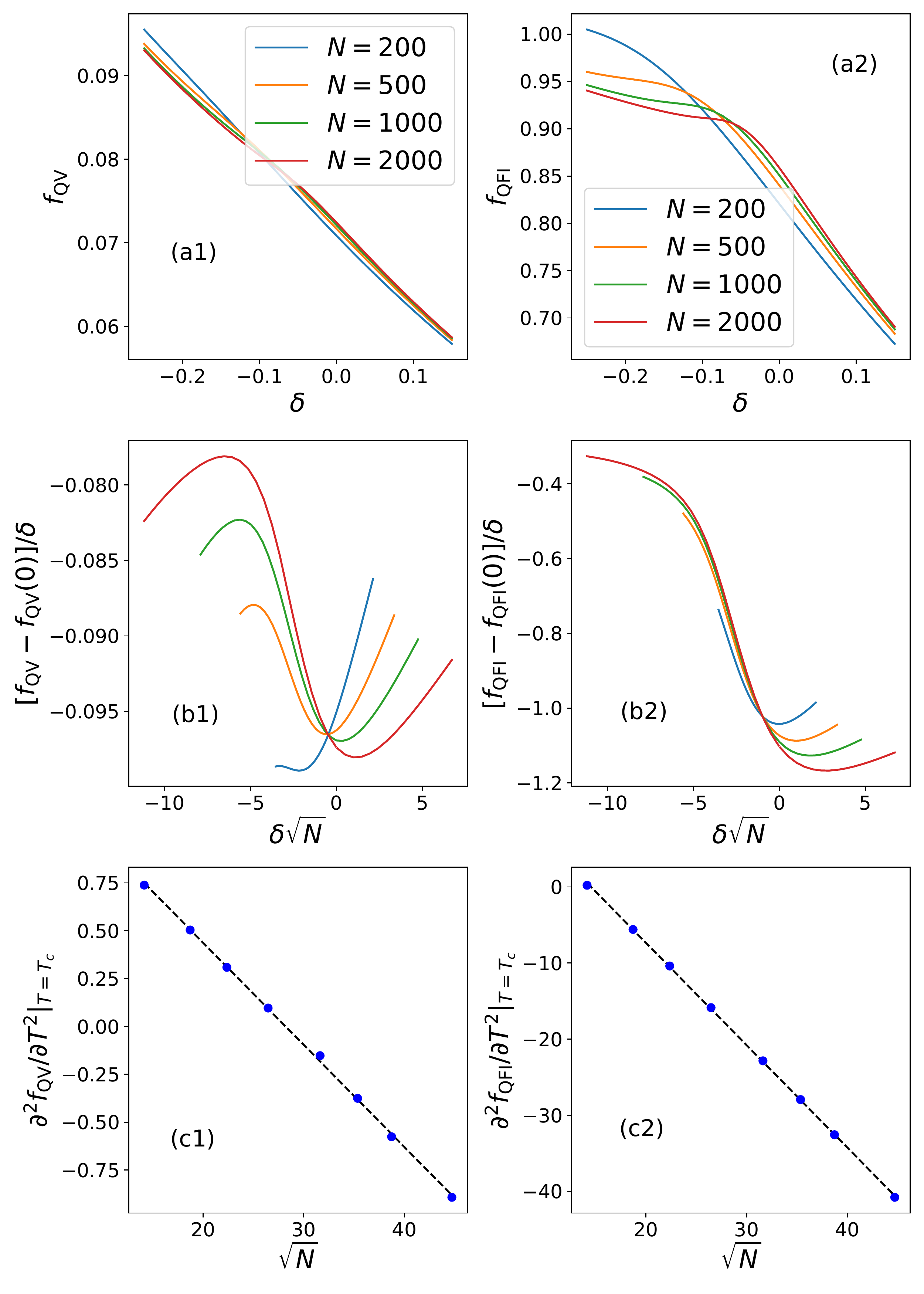}
		\caption{Weak singularity of the QV and QFI around the thermal critical point of the TFIM with infinite-range interactions ($g=0.8. T_c\approx 0.3641 J$). (a1) QV per spin  $f_{\rm QV} = N^{-1} {\rm Var}_Q(J^z)$ as a function of the reduced temperature $\delta = T/T_c - 1$ for various system sizes ($N=200, 500, 1000, 2000$); (b1) same for the QFI. (b1)-(b2) Linearized first derivative, showing a discontinuity building up with system size; (c1)-(c2) second derivative at the critical point, exhibiting a $\sqrt{N}$ scaling.}
		\label{fig_QV_Ising_all_to_all}
	\end{figure}

\textit{Critical behavior of quantum fluctuations.}
Fig.~\ref{fig_QV_Ising_all_to_all} shows the QV (panel a1) and QFI (a2) per spin across the thermal phase transition [respectively $f_{\rm QV} = N^{-1} {\rm Var}_Q(J^z)$ and $f_{\rm QFI} = N^{-1} F_Q(J^z)$], clearly exhibiting the buildup of a singularity for $N \to \infty$. 
 
According to the discussion in the main text, the dominant contribution to the singularity is of the form $[f_{\rm QV}]_s, [f_{\rm QFI}]_s \sim |\delta|^{2\nu z - \gamma}$ with $\delta = (T - T_c)/T_c$. In the present case ($\nu=1/2, z=1, \gamma=1$), we have $2\nu z - \gamma=0$. Since a jump singularity (compatible with a 0 exponent) is not observed in the QV and QFI [Fig.~\ref{fig_QV_Ising_all_to_all}(a1-a2)], the critical singularity must come from the second term in Eq.~3 of the main text, that is: $[f_{\rm QV}]_s, [f_{\rm QFI}]_s \sim |\delta|^{4\nu z - \gamma}=|\delta|$. We expect therefore a discontinuous change of slope of the QV and QFI at $T=T_c$. More precisely, a finite-size scaling analysis shows that the first derivative of the QV and QFI with respect to the temperature admits the scaling form $ \partial_T [f_{\rm QV}]_s,  \partial_T [f_{\rm QFI}]_s \sim f(\delta \sqrt N)$. Fig.~\ref{fig_QV_Ising_all_to_all}(b1-b2) indeed confirms this scaling behavior, although the finite-size effects are still important for the largest sizes we investigate ($N=2000$), leading to an imperfect data collapse. Finally, this scaling form implies that the second derivative at the critical point diverges as $\sqrt N$, as confirmed on Fig. \ref{fig_QV_Ising_all_to_all}(c1-c2).

	In summary, our numerical and theoretical findings are consistent with a dynamical exponent $z=1$ for the infinite-range transverse-field Ising model [Hamiltonian in Eq. \eqref{eq_Ising_LMG}] at the finite-temperature critical point, responsible for a weak singularity of both the QV and QFI of the form $|T-T_c|$ in the thermodynamic limit. The likely explanation for the absence of such a singularity in the QFI as computed in Ref.~\cite{Haukeetal2016} is an incorrect account for the sectors $J < N/2$ in Eq.~\eqref{eq_S_omega_Ising_all_to_all}. While the ground state only lies in the $J=N/2$ sector, all sectors with their exponential degeneracy $D(J)$ contribute to the thermal properties.

	\begin{figure*}[ht!]
	\centering
		\includegraphics[width=0.7\linewidth]{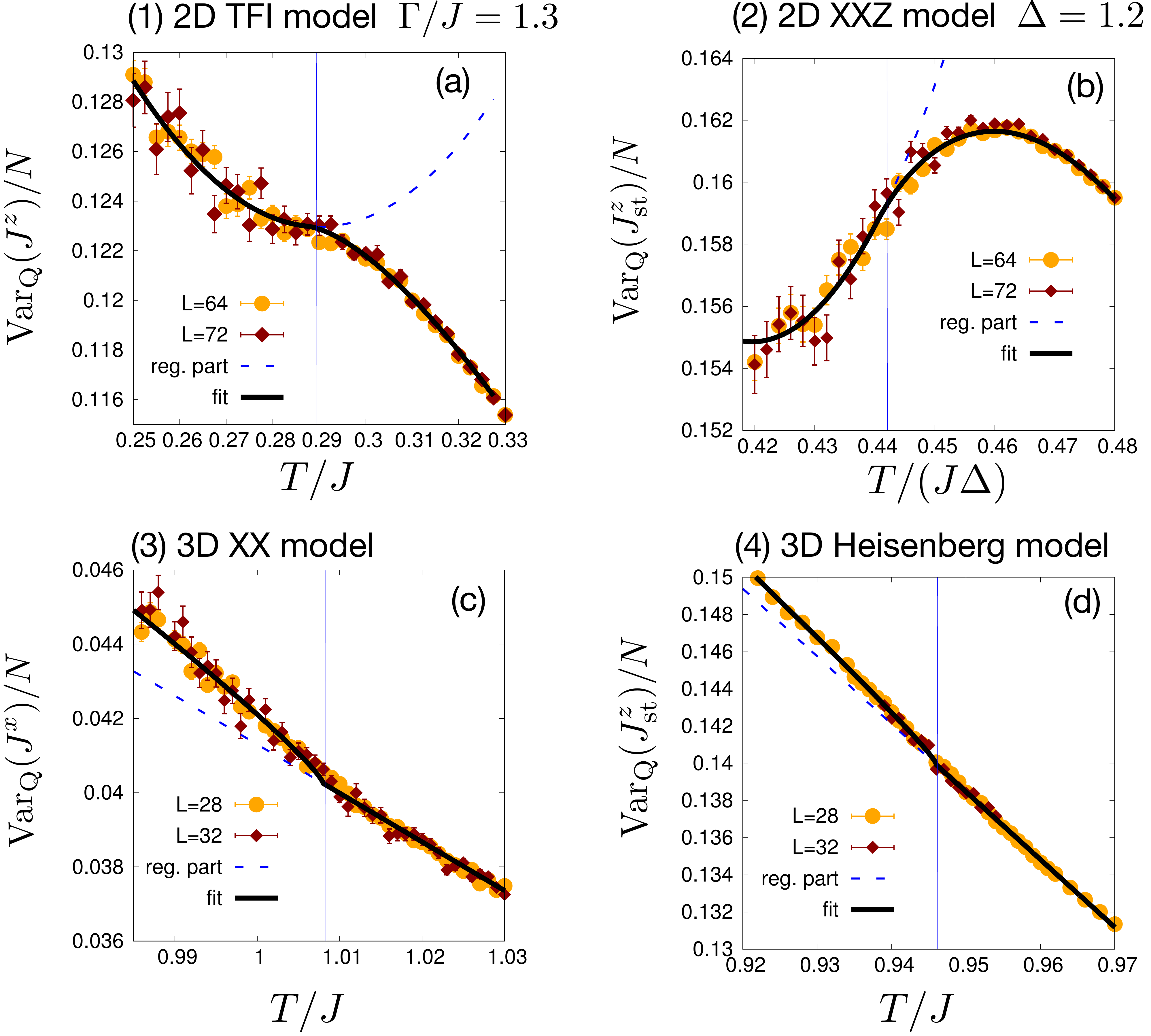}
		\caption{Singularity of the quantum variance in the vicinity of the transition. The four panels refer to the four models discussed in the main text, and referenced to in the panel labels. In all panels the points are QMC data for two different $L^d$ lattices, the black line is an example of a fit to the function given in the main text, and the dashed line shows the regular part. The vertical blue line indicates the position of the transition.}
		\label{Qv_zoom}
	\end{figure*}

\section{Discussion of the Quantum Monte Carlo results and their fits}

\emph{Quantum Monte Carlo results.}
Fig.~\ref{Qv_zoom} shows a zoom of the quantum variance of the order parameter close to the thermal phase transition for the four models studied via QMC -- compare Fig. 1 of the main text for an extended view. In each figure two system sizes are compared ($L=64$ and $L=72$ for the 2$d$ models; $L=28$ and $L=32$ for the 3$d$ models), showing that size effects are very small for the system sizes considered; and that the numerical precision on the quantum variance degrades with system size, especially below the phase transition, because of the divergence of the two quantities whose difference defines the quantum variance. The figure also shows the fitting function, visually highlighting the consistency between the singularity contained in the fitting function and the (weakly) singular behavior exhibited by the quantum variance.  

In particular, in order to achieve the numerical precision on the quantum variance that allowed us to perform meaningful fits, we had to resort systematically to choosing a computational basis \emph{orthogonal} to the order parameter, so that the quantum fluctuations of the order parameter are estimated as off-diagonal observables. This aspect is motivated by the fact that in Stochastic Series Expansion, as well as in any finite-temperature quantum Monte Carlo schemes, the estimate of the off-diagonal correlations is made during the (directed-loop or worm) update \cite{SyljuasenS2002}, and it therefore enjoys a much richer statistics than diagonal observables which are sampled instead only in between updates.

\emph{Alternative fitting strategies for the two-dimensional models.} As stated in the main text, a fundamental difficulty in fitting the QV data with a singular and a regular part is that both parts are non-diverging, so that discriminating between them is generally non-trivial. For the fits of the QMC data presented in the main text, we have chosen the regular part to fully account for the QV on one side of the transition; while the singular part adds up to the regular part on the opposite side. This choice is justified in the case of the 3$d$ models we considered, where one clearly observes a slope discontinuity on the two sides of the transition (Fig.~\ref{Qv_zoom}(c-d)), which can be accounted for by a linear (and regular) QV above the transition and a weakly non-linear QV below; we have not found an alternative hypothesis that could lead to good fits.

\begin{figure*}[ht!]
	\centering
		\includegraphics[width=0.7\linewidth]{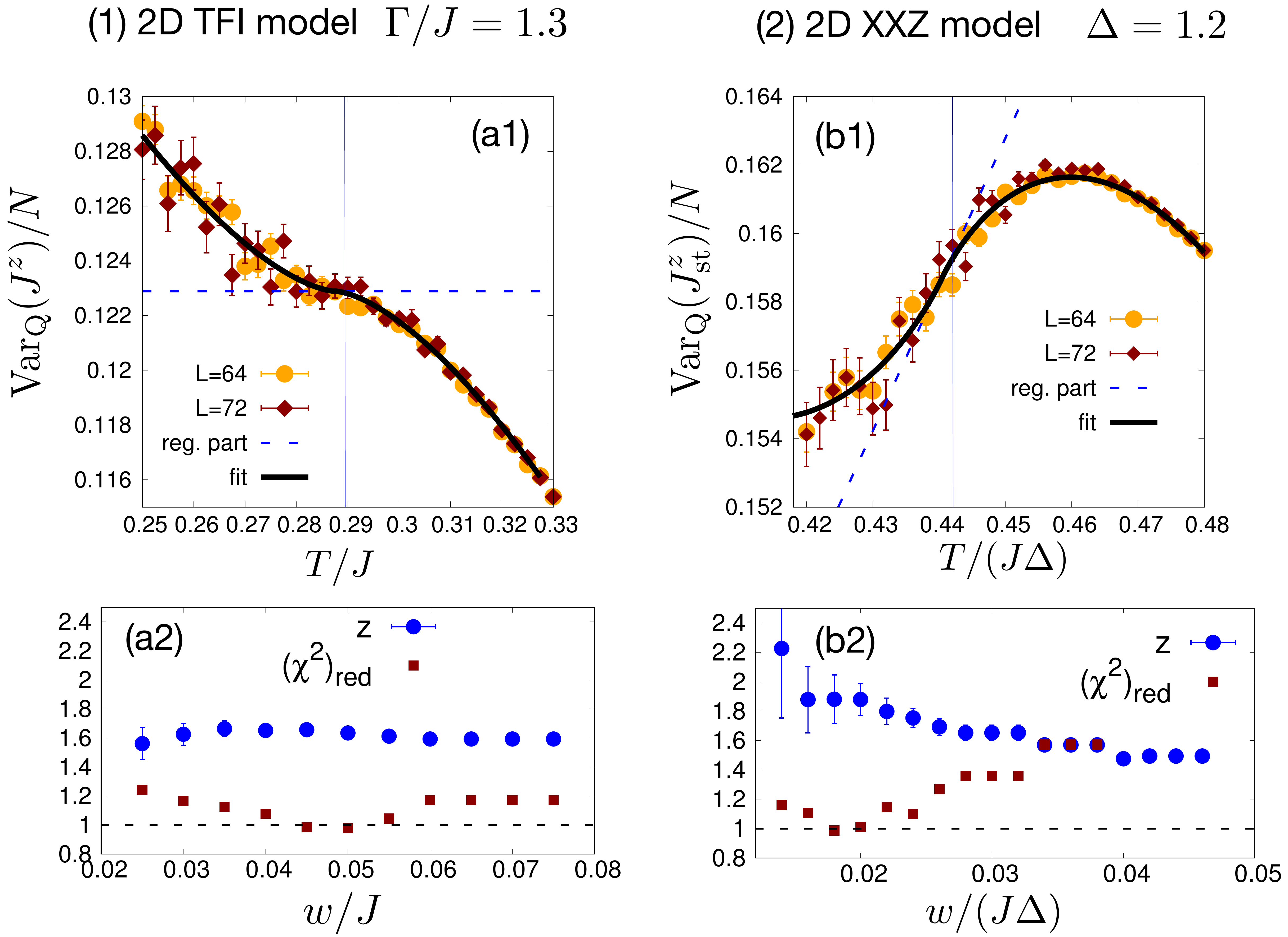}
		\caption{Results of the alternative fitting scheme for the 2$d$ TFI and XXZ models. All symbols in panels (a1) and (b1) are as in Fig.~\ref{Qv_zoom}, except that the solid black curve now indicates the fit to the modified fitting function (see text). Panels (a2) and (b2) show the fitted $z$ exponent and the reduced $\chi^2$ of the fit as a function of the fitting window. In the case of the 2$d$ XXZ model, the fits have been performed on an asymmetric window $[T-w, T+1.5*w]$ due to the higher quality of the data above the transition.}
		\label{altfits}
	\end{figure*}

On the other hand, for the 2$d$ models, the curvature of the QV captured by the regular part on one side of the transition could as well be accounted for by the singular part being finite on both sides. Therefore we have analyzed the 2$d$ data with an alternative hypothesis: 1) for the 2$d$ TFI model, we have chosen to take the regular part as a constant, and to allow for a singular part on both sides of the transition, fitting therefore with the parameters $a_0, A_+, A_-, z$; 2) for the 2$d$ XXZ model, we have also turned on a linear term (with coefficient $a_1$) in the regular part, due to the manifestly finite slope at the critical point. The resulting fits are shown in Fig.~\ref{altfits}, exhibiting a quality which is fully comparable to that of the fits shown in the main text. In the case of the 2$d$ XXZ model, the $z$ exponent resulting from the alternative fitting scheme, $z = 1.88(10)$, is compatible with the one we have obtained in the main text ($z=1.95(10)$), so that the estimate of the $z$ exponent appears to be robust. As for the 2$d$ TFI model, on the other hand, the alternative fitting scheme gives systematically lower values for the $z$ exponent, with a final estimate -- drawn using the same criteria as the one declared in the main text -- of $1.65(5)$. This is to be compared to $z=1.95(15)$ obtained with the fitting procedure of the main text. Interestingly, the values of $z$ obtained with the alternative fitting scheme for both models (2$d$ quantum Ising and XXZ) are closer to the experimental values obtained by various methods on $2d$ quantum magnets, see Refs.~\cite{Buccietal1974, Hutchingsetal1982, Slivkaetal1984, tsengetal2016,thesis_tseng}. Our current conclusion is that theoretical data of higher quality, or a fundamental justification on the choice of the fitting function, would be needed to discriminate between the two fitting schemes, and draw a conclusive value of $z$ for the model in question.             

\end{document}